# Spoken Digit Recognition and Speaker Classification by Nonlinear Interfered Spin Wave-Based Physical Reservoir Computing

Sota Hikasa[1,2], Wataru Namiki[1], Daiki Nishioka[3], Maki Nishimura[1,2], Ryo Iguchi[1], Kazuya Terabe[1] and Takashi Tsuchiya[1,2*]

[1] Research Center for Materials Nanoarchitectonics (MANA), National Institute for Materials Science (NIMS), 1-1 Namiki, Tsukuba, Ibaraki, 305-0044, Japan.

[2] Faculty of Advanced Engineering, Tokyo University of Science, 6-3-1 Niijuku, Katsushika, Tokyo 125-8585, Japan.

[3] International Center for Young Scientists (ICYS), NIMS, 1-1 Namiki, Tsukuba, Ibaraki, 305-0044, Japan.

*Email: TSUCHIYA.Takashi@nims.go.jp

Abstract

Recently, artificial-intelligence (AI) technologies have been increasingly utilized in a wide range of real-world applications. Speech recognition is one of these practical AI tasks and is regarded as a key application for edge AI systems. Consequently, speech recognition has been widely employed as a representative benchmark task for assessing the performance of physical reservoir computing (PRC). Although many PRCs have performed this task, the majority of them rely on the frequency-extraction preprocessing method, such as a cochleagram and mel-frequency cepstrum. Especially about the cochleagram, this method enables high-accuracy recognition; however, it requires a substantial computational cost for preprocessing and is unsuitable for edge computing, due to the limited resources. In this study, we employed a nonlinear interfered spin wave-based PRC, which demonstrated superior computational performance in mathematical tasks. Using this PRC, we evaluated the performance for two types of speech recognition, spoken digit recognition and speaker classification under four configurations: cochleagram-alone, interfered spin wave-based PRC with cochleagram, baseline without PRC, and interfered spin wave-based PRC alone to quantify the contributions of the cochleagram and of the interfered spin wave-based PRC for each task. As a result, although the cochleagram alone yielded accuracies around 90 % for both tasks, the accuracy reached 85.8 % for speaker classification when only the interfered spin wave-based PRC was used. These results indicate the potential of the proposed PRC to handle speech recognition tasks without cochleagram preprocessing.

## I. Introduction

Physical reservoir computing (PRC), which exploits the inherent nonlinear dynamics of physical systems (e.g., optical circuits,[1-6] ion-gating reservoirs,[7-19] spintronics devices,[20-36] memristors,[37-42] among others.[43-52]), has emerged as a promising approach for edge AI (artificial-intelligence) owing to its low power consumption and high-speed operation.[52-55] Interfered spin waves are one of the most promising candidates because they possess the three characteristics required for physical systems: nonlinearity, short-term memory, and response diversity, which enable a nonlinear transformation of the input data into a high-dimensional space for time-series processing. First, strong nonlinearity originates from spin-wave propagation and interference. Second, short-term memory effects arise from the relaxation dynamics of the spin waves. Third, the use of many detectors provides a high-dimensional space. By virtue of these properties, previous experimental demonstrations of spin-wave interference have reported superior computational performance on nonlinear autoregressive moving-average (NARMA) benchmark tasks.[27] However, many of the currently reported spin wave-based PRCs are limited to demonstrations in mathematical tasks such as NARMA, and chaotic time-series prediction, and their processing performance for practical tasks remains unclear.[27-32]

Speech recognition is one of the representative tasks in modern AI-enabled devices. Especially in edge-AI systems, speech recognition is increasingly required to be performed directly on devices without relying on cloud computing from the viewpoint of privacy protection and latency. For example, tasks such as speaker identification using speech signals benefit from on-device processing without transmitting voice data to external servers. Accordingly, it has frequently been adopted as a benchmark task in studies on PRCs.[5-6,26,34,36,40-42,45-51] In contrast to the mathematical benchmarks (e.g. NARMA), speech recognition is a practical time-series classification task, making it well suited for assessing the capability of PRCs to process real-world temporal data. In speech recognition, discriminative features for classification are derived not only from the dynamics of signal intensity, but also from frequency components and their temporal distributions, which reflect speaker-specific anatomical structures and speaking habits. Consequently, sufficient feature-extraction capability of the physical system is a crucial requirement for high recognition performance. In most previous studies, feature extraction is not achieved solely by the reservoir itself but is instead supported by preprocessing techniques such as cochleagram.[56] However, although this preprocessing method markedly improves accuracy, it has been pointed out that it obscures the evaluation of the intrinsic computational capability of PRCs in benchmark tasks.[51] Simultaneously, achieving high recognition performance without such preprocessing remains challenging. For edge-AI applications, PRCs should operate under severe constraints on computational resources and power consumption, while also benefiting from low latency due to minimal cloud communication. From these perspectives, preprocessing methods of cochleagram are often unsuitable because they add power consumption and latency. Various approaches have been proposed, yet they still fall short of the required performance.[46-47,50-51]

Therefore, in this study, the computational performance of the nonlinear interfered spin wave-based PRC in speech recognition tasks by considering the contribution from preprocessing to explore its potential for practical applications. The two types of speech recognition tasks were considered: spoken digit recognition, which aims to identify spoken digits, and speaker classification, which aims to identify the speaker. These tasks were performed under four conditions combining the presence or absence of the cochleagram and interfered spin wave-based PRC in order to examine the roles of each component in computational performance. The results confirm that the frequency decomposition capability provided by the cochleagram significantly contributes to the computational performance in both speech recognition tasks. On the other hand, while the accuracy remained low for the spoken digit recognition task, high classification performance was obtained in the speaker classification task when interfered spin wave-based PRC was employed alone. These results indicate that even within speech recognition, the information processing capabilities required of physical systems for PRC differ depending on the specific task. These results suggest that the nonlinear interfered spin wave-based PRC has the potential to handle speech recognition, a key task in edge-AI systems, without relying on cochleagram-based frequency decomposition.

## II. Methods

### A. Device design and measurement setup for interfered spin wave-based PRC

PRC is a machine learning framework suitable for processing time-series data such as chaotic time series, electrocardiogram (ECG) signals, and speech signals.[57] In PRC, tasks such as waveform prediction, anomaly detection, and classification can be performed by exploiting nonlinear and hysteretic dynamical responses in the reservoir layer (Fig. 1(a)). In this study, we employed spin-wave interference as the physical phenomenon that realizes the nonlinear dynamics required for PRC. As the spin-wave medium, a double-sided polished yttrium iron garnet ($Y_3Fe_5O_{12}$: YIG) single crystal, which is a ferrimagnet, with 111-oriented surfaces and provided by TwoLeads Corporation, was used. YIG is widely used in spin-wave-based devices because it exhibits a low Gilbert damping constant and a long spin-wave propagation length.[58-59] The size of the YIG sample is 7 mm × 7 mm × 0.5 mm. Figure 1(b) shows the antenna structure for exciting spin waves. A total of ten coplanar waveguides (CPWs) were fabricated on the YIG single crystal using laser lithography techniques followed by electron-beam evaporation. Among ten CPWs, two CPWs serve as exciters (Ext.) and other seven as detectors (Det.). By using two exciters, the excited spin waves nonlinearly propagate and interfere with each other, resulting in the nonlinear transformation of the input data, which can be captured by the seven detectors more than the previous ones.[27-28,32] Each antenna consists of one single line (width: 10 μm) and two ground lines (width: 20 μm), where the total width of CPWs is 70 μm, formed from Ti (10 nm)/Au (90 nm) thin films. The antenna distance between each antenna is shown in Fig. 1(c) and was designed to be asymmetric to achieve the diverse output waveforms.

Thanks to this asymmetric structure, information spread within the YIG medium due to propagation and interference can be acquired more efficiently. The experimental setup for the interfered spin wave-based PRC is shown in Fig. 1(d). A magnetic field was applied perpendicular to the YIG surface, thus exciting the magnetostatic forward-volume mode spin waves. While the two exciters were connected to an arbitrary waveform generator (AWG; Keysight M8195A), the seven detectors (from Det. 1 to Det. 7) were connected to an oscilloscope (Tektronics MSO68B) via rf probers in the speech recognition task. Therefore, the number of detectors $N_{\mathrm{det.}}$ was 7. The input data were generated using the AWG with an output amplitude set to 560 mVpp and a sampling rate of 16 GS/s and amplified by a rf amplifier (Mini-Circuits ZVA-0.5W303G) before being fed into the YIG medium. The spin-wave output signal was also amplified using another rf amplifier (Pasternack PE15A3269) and measured with the oscilloscope. The output signals were averaged over 500 repetitions to improve the signal-to-noise ratio.

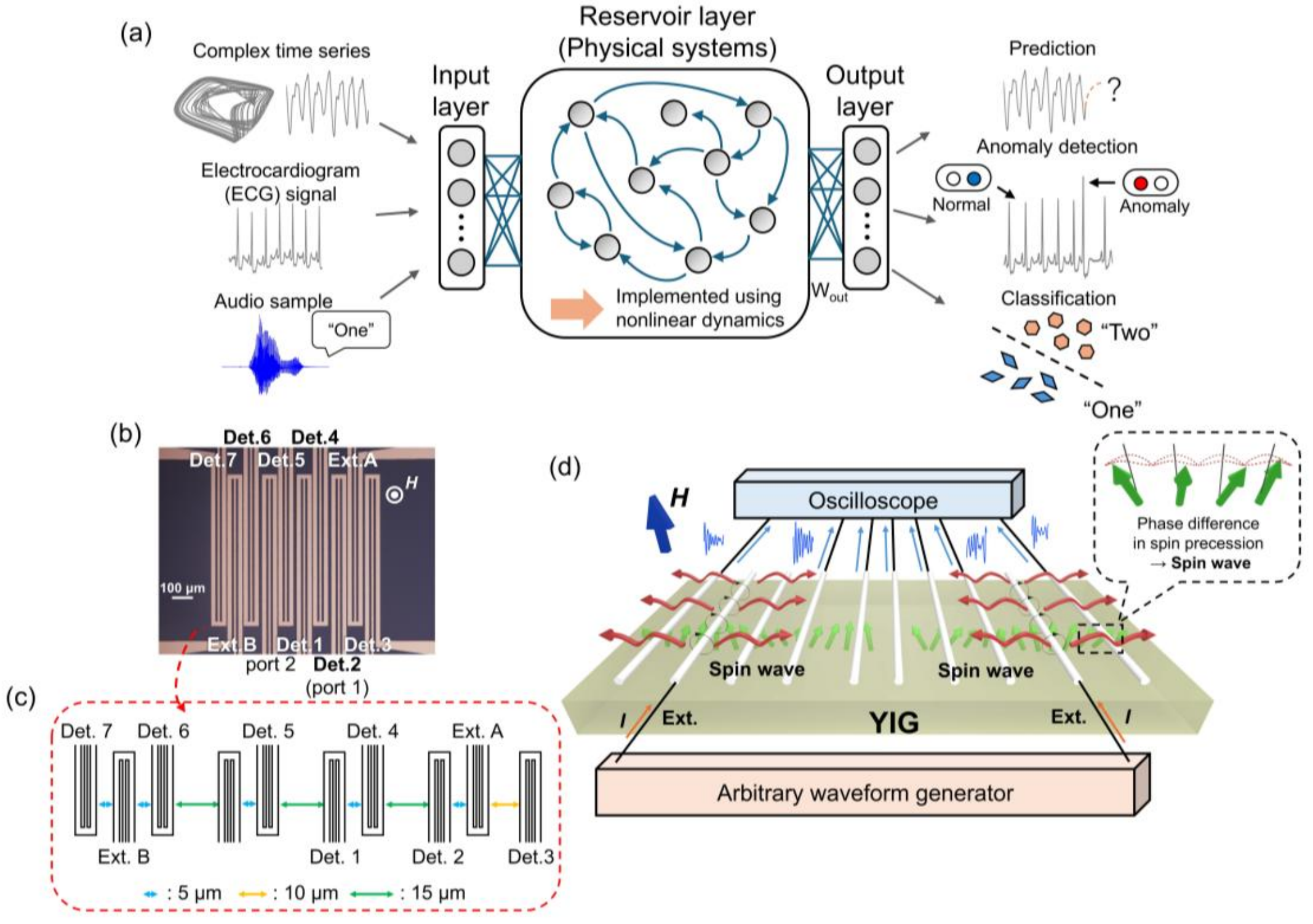

FIG.1 (a) Schematic diagram of PRC. When time-series data are input into the reservoir layer, features are extracted through the dynamical response of the reservoir, enabling tasks such as classification. (b) Optical-microscope image of the antenna fabricated on the YIG single crystal. Each antenna is CPW and the total width of the CPW is 70 μm. (c) The antenna layout and distances between each CPWs. (d) Schematic illustration of the experimental setup used in this study.

B. Reservoir computing for speech recognition with and without preprocessing

The audio dataset for the input data of each speech recognition task is the TI-46 speech corpus. This data set contains 500 audio files and consists of utterances of the digits zero–nine spoken ten times each by five female speakers.[60] The reservoir output is obtained by multiplying the reservoir-state matrix $\boldsymbol{X}$ with the output-weight matrix $\boldsymbol{W}$, as shown below.

$$\boldsymbol{Y} = \boldsymbol{W} \cdot \boldsymbol{X} \tag{1}$$

Here, $\boldsymbol{Y} = [\boldsymbol{y}(1), \boldsymbol{y}(2), \cdots, \boldsymbol{y}(k), \cdots, \boldsymbol{y}(T)]$ is the reservoir-output matrix, where $\boldsymbol{y}(k)$ is reservoir-output vector, $k$ represents the time index at which the reservoir output is evaluated, and $T$ is the number of time index for training or testing phase. $\boldsymbol{W} = \left[\boldsymbol{b}, \boldsymbol{w}_1, \boldsymbol{w}_2, \cdots, \boldsymbol{w}_{N_{\mathrm{tot}}}\right]$ is the output-weight matrix, where $\boldsymbol{b}$ represents the bias vector, and $N$ is the total number of nodes. $\boldsymbol{X} = [\boldsymbol{x}(1), \boldsymbol{x}(2), \cdots, \boldsymbol{x}(k), \cdots, \boldsymbol{x}(T)]$ is the reservoir-state matrix, where $\boldsymbol{x}(k) = [1, X_1(k), X_2(k), \cdots X_N(k)]$ represents the reservoir-state vector. The reservoir-state matrix is represented by the total number of nodes and the total number of time indices. Using this equation, the reservoir-output matrix is first computed. Then, for each time index $k$, the label corresponding to the maximum response in the reservoir output is selected as the output at that time index. After repeating this procedure over all time indices, a single output label corresponding to one audio data is determined by applying a winner-take-all method (i.e., by selecting the most frequent label across all time indices for the same audio sample). This label is taken as the final reservoir response. The output-weight matrix was optimized using linear regression, as follows:

$$\boldsymbol{W} = \boldsymbol{D}\boldsymbol{X}^{\mathrm{T}}(\boldsymbol{X}\boldsymbol{X}^{\mathrm{T}})^{-1} \tag{2}$$

where $\boldsymbol{D} = [\boldsymbol{d}(1), \boldsymbol{d}(2), \cdots, \boldsymbol{d}(k), \cdots, \boldsymbol{d}(T)]$ is the target matrix, and $\boldsymbol{d}(k)$ is the one-hot-vector at each time index. A 10-fold cross-validation scheme was employed to achieve more reliable weight optimization and performance evaluation because of the limited amount of data. In each fold, 450 of the total 500 audio data were used for training phase, while the remaining 50 audio data were used for testing phase. In this study, accuracy, defined as the ratio of the number of correctly classified data to the total number of data, was used as the performance evaluation index.

In this study, to evaluate the influence of the cochleagram on each task and the feature extraction capability of the nonlinear interfered spin wave-based PRC, we constructed four types of reservoir states: (i) cochleagram-alone, (ii) cochleagram+PRC, (iii) w/o cochleagram and PRC, and (iv) PRC-alone. The comparison of processing is shown in Fig. 2. For all cases utilizing the PRC (case (ii) and (iv)), the applied magnetic field was set to the 220 mT.

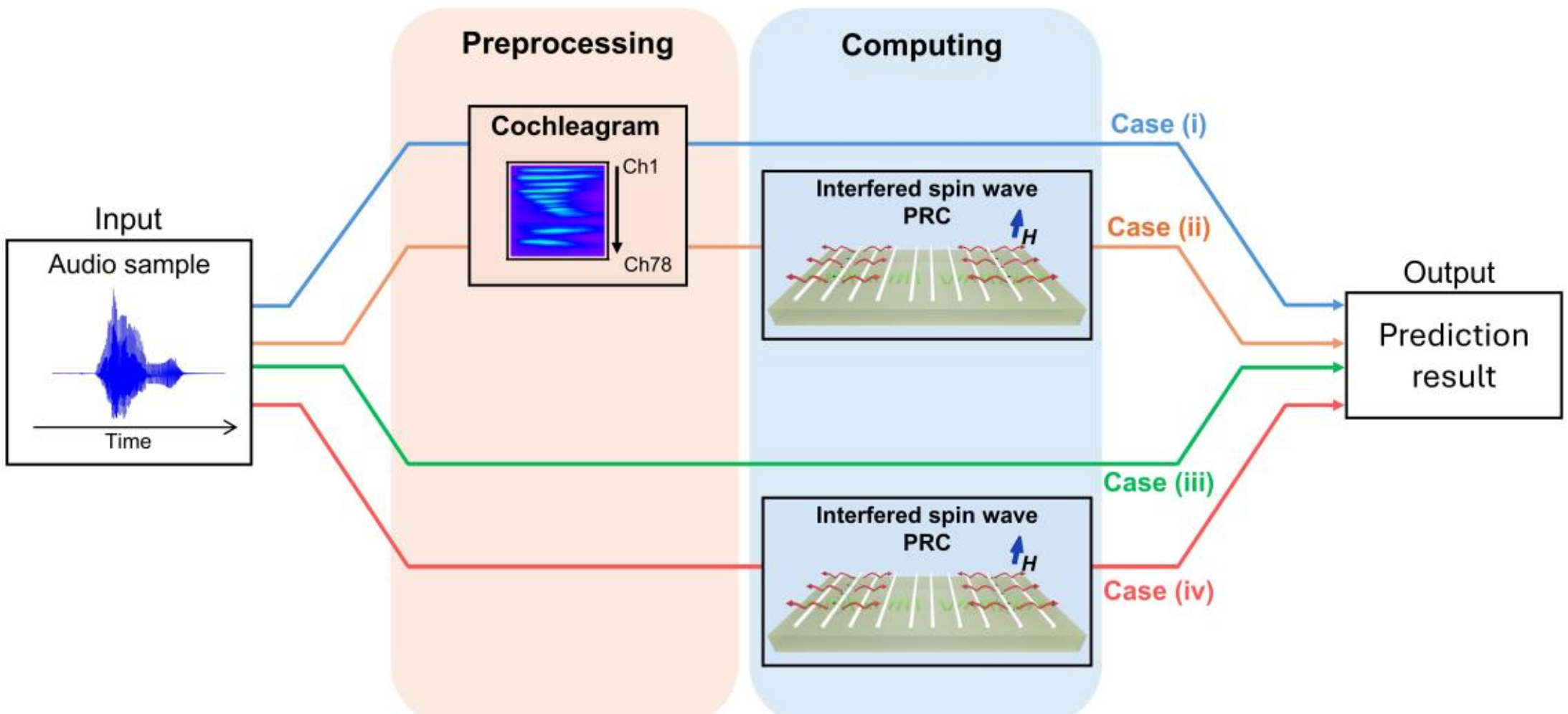

FIG. 2 Comparison of the four configurations used in this study. Case (i), shown in blue, represents the cochleagram-alone condition, where cochleagram processing was applied in the preprocessing phase while no processing was performed in the computing phase. Case (ii), shown in orange, represents the cochleagram+PRC condition, where cochleagram processing was applied in the preprocessing phase, and this processed-data was fed into the interfered spin wave-based PRC. Case (iii), shown in green, represents w/o cochleagram and PRC condition, where neither cochleagram preprocessing nor the interfered spin wave-based PRC was used, and only pulse conversion and pulse peak extraction were performed. In other words, this corresponds to case (iv) without the contribution of the interfered spin wave-based PRC. Case (iv), shown in red, represents the PRC-alone condition, where the input signal was only converted into pulse sequences and fed into the interfered spin wave-based PRC.

In cases using cochleagram corresponding to case (i) and (ii), the evaluation flowchart is shown in Fig. 3(a). In the preprocessing step, the audio data were segmented into the time domain, and each segment was transformed into 78 frequency-specific components ($N_{\mathrm{ch}}$), resulting in cochleagram representation. The number of time domains for each audio data is up to 92. Case (i) directly used this cochleagram for creating reservoir states and evaluation. Whereas in case (ii), this cochleagram-processed data was normalized to the range from 0.1 to 1. Then, for each frequency channel, the time-series data were converted into a pulse sequence with 15 ns pulse interval (the minimal preprocessing needed to excite spin waves) and fed into the YIG medium via AWG.[27] In case (ii), the time index corresponds to a single pulse input, and the total number of time indices is 22265. 18 virtual nodes ($N_{\mathrm{v}}$) were extracted per detector from each output waveform corresponding to a single input for both tasks. As a result, the total number of nodes $N$ was 9828 (= 18 ($N_{\mathrm{v}}$) × 7 ($N_{\mathrm{det}}$) × 78 ($N_{\mathrm{ch}}$)), and the reservoir-state matrix was given by $\boldsymbol{X} \in \mathbb{R}^{9829\times 22265}$, which includes the bias term. The optimized readout weights were then obtained, and the reservoir output was generated using Eqs. (1) and (2), respectively.

In cases not using cochleagram corresponding to case (iii) and (iv), the evaluation flowchart is shown in Fig. 3(b). The five speakers were labeled S1 to S5. In these cases, the audio data were only converted into a pulse sequence with an interval of 15 ns. To construct the reservoir-state matrix, the time-series data corresponding to each audio sample were divided into several time-domain windows, each containing a fixed number of input pulses. The virtual nodes extracted from each window were grouped into a single time index, and prediction labels were calculated for each time index. To ensure consistent segmentation, all speech signals were trimmed so that the starting points were aligned across samples. The end points were chosen after the speech signal had sufficiently decayed and were adjusted so that the total number of input pulses was divisible by the number of input pulses per time-domain window. The number of time-domain windows is determined by dividing the number of peaks in each trimmed audio sample by the number of input pulses per window. In case (iii), the peak values of the input-pulse voltage were used as the time-series data. These data were segmented into time-domain windows consisting of 1000 input pulses, resulting in a reservoir-state matrix of $\boldsymbol{X} \in \mathbb{R}^{1001\times3946}$, which includes the bias term. On the other hand, in case (iv), 30 virtual nodes ($N_{\mathrm{v}}$) were extracted from the obtained spin-wave responses corresponding to a single input pulse which were divided into several time-domain windows containing 90 pulse responses ($N_{\mathrm{pulse}}$). Therefore, the total number of nodes $N$ became 18900 (= 90 ($N_{\mathrm{pulse}}$) × 30 ($N_{\mathrm{v}}$) × 7 ($N_{\mathrm{det}}$)), and finally the size of reservoir-state matrix was $\boldsymbol{X} \in \mathbb{R}^{18901\times41235}$, which includes the bias term. After extraction, the optimized readout weights were applied to generate one reservoir output per window. Consequently, the final classification for the audio sample was then determined by majority voting among these outputs.

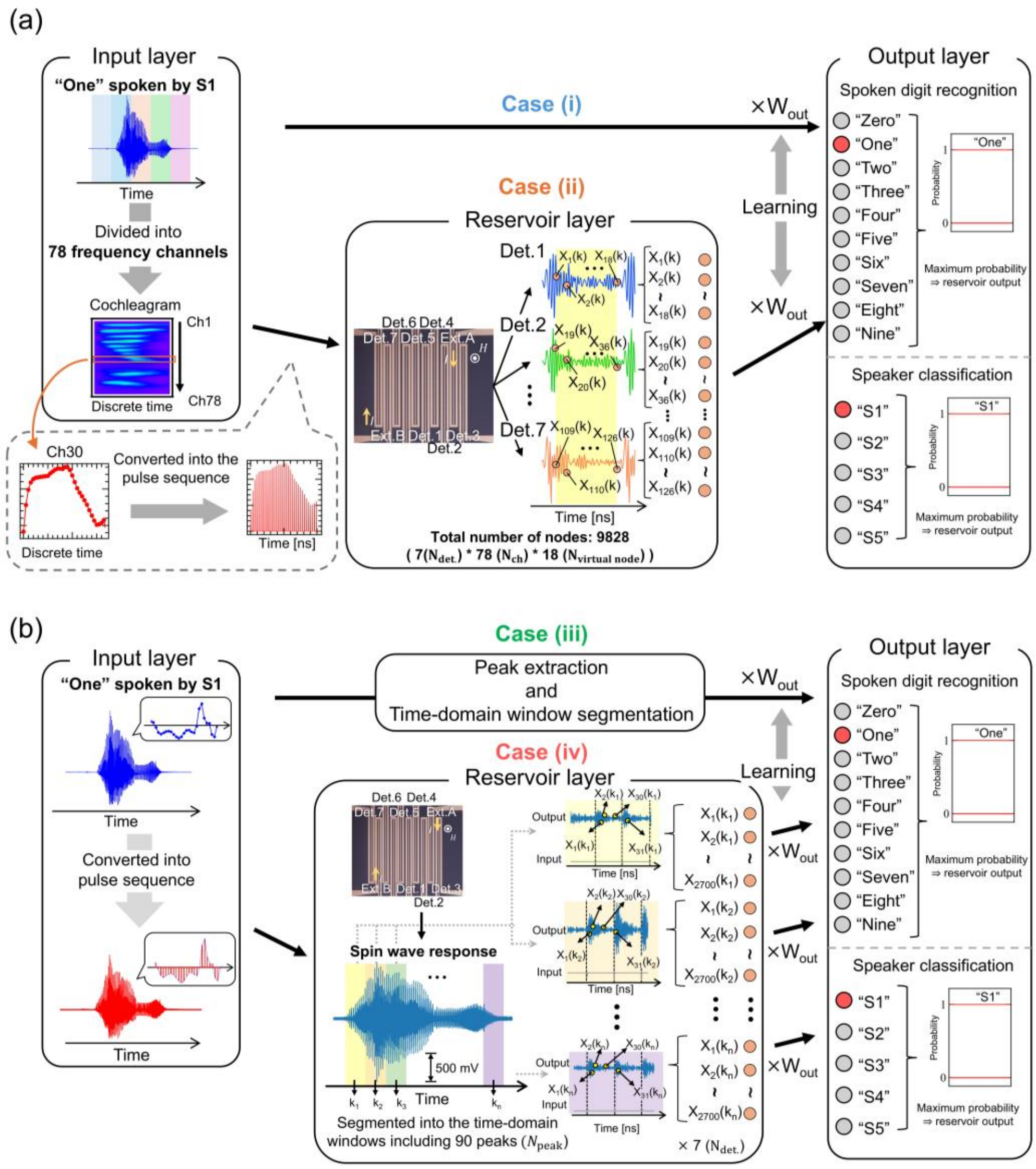

FIG.3 The schematic diagrams of processing for each case. (a) The cases using cochleagram. In case (i), cochleagram-preprocessed data were used for both tasks. On the other hand, in case (ii) the preprocessed data were then converted into a pulse sequence with a 15 ns interval and fed into the YIG medium. (b) The cases not using cochleagram. The audio data were converted into the pulse sequence with the 15 ns interval. In the case (iii), the converted data were directly used for performing speech recognition tasks. Whereas, in the case (iv), the converted data were fed into YIG medium.

## III. Results and discussion

### A. Magnetic properties of the YIG single crystal

To investigate the dependence of spin-wave excitation on an external magnetic field in the YIG single crystal, the transmission spectroscopy measurements were performed using a vector network analyzer (VNA; Copper Mountain Technologies S5085). Ports 1 and 2 in Fig. 1(b) were connected to the VNA (port 1 corresponds to Det. 2) for measuring the transmission S-parameter ($S_{21}$).

Figure 4(a) shows the de-embedded $S_{21}$, for which the reference data was measured at 0 mT, revealing how the magnetic field affects both the intensity and the frequency of the excited spin waves. It shows that both the intensity and frequency of the excited spin waves depend on the magnetic field. The magnetic-field dependence of this spin-wave response was also observed in oscilloscope measurements, as shown in Fig. 4(b). Here, the excited spin wave was described by the equation as follows:

$$f(H) = \gamma \mu_0 \sqrt{(H - H_\mathrm{a})\left\{(H - H_\mathrm{a}) + M\left(1 - \frac{1 - e^{-kd}}{kd}\right)\right\}} \tag{3}$$

where $f(H), H, \gamma, \mu_0, H_\mathrm{a}, M, k$, and $d$ are frequency, perpendicularly applied magnetic field, gyromagnetic ratio, permeability of free space, magnetic anisotropy, saturation magnetization, wavenumber of the excited spin wave, and the thickness of YIG sample, respectively.

Here, $\gamma$ is 28 $\mathrm{MHz \cdot mT^{-1}}$, and the wave number $k$, whose values and calculation methods are described in Ref 31. From the fitting result of this measurement, saturation magnetization $\mu_0 M$ and magnetic anisotropy $\mu_0 H_\mathrm{a}$ were (191.6 $\pm$ 7.8) mT and (158.9 $\pm$ 8.2) mT, respectively.

In this study, two exciters (Ext. A and Ext. B) were used to induce spin wave interference. Owing to the magnetic dipole interaction, this nonlinear interference is enhanced, and its magnitude further increases with the spin-wave amplitude. Figures 4(c)–(d) show that the difference between the actual interference signal and the linear sum of the two individual excitations (Ext. A and Ext. B) yields a wave-packet-like waveform. This result indicates that the interference observed in our experiments is nonlinear, as reported in the previous research.[27] Input data undergoes complex transformation through spin wave propagation and nonlinear interference and is mapped in a spatially distributed manner.[19,27-28,31] Figure 4(e) shows the output waveforms detected at Det. 1, 3, 5, and 7 (output waveforms observed at the other detectors are shown in Fig. 6, Appendix A), and these outputs represent diversity. These diverse output waveforms demonstrate that the nonlinear, spatially distributed transformation of the input can be efficiently captured by multi-detection. Such diversity provides richer information, thereby improving computational performance.

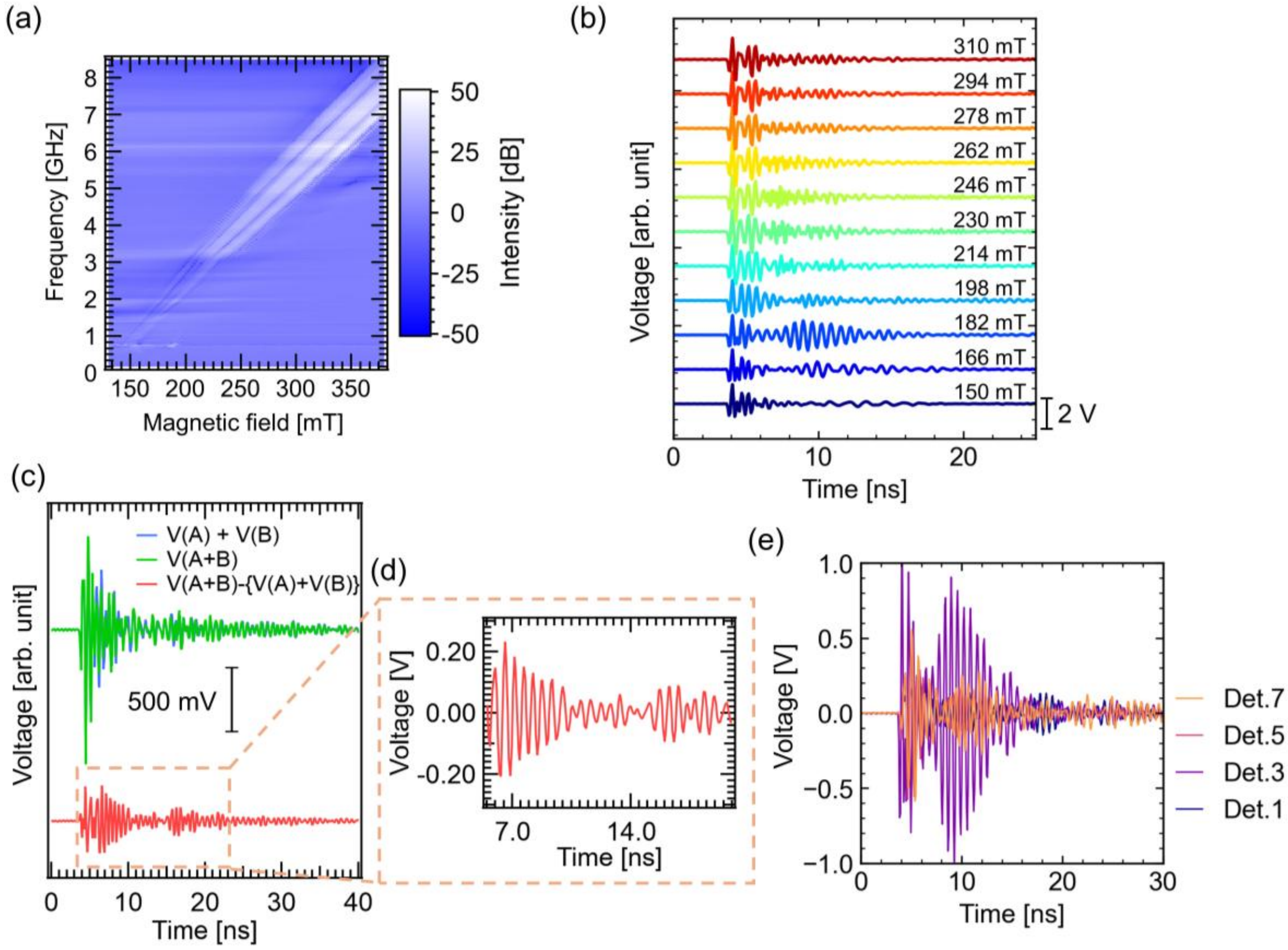

FIG. 4 (a) Spin wave spectroscopy of de-embedded transmission signals. Port 1 and port 2, shown in Fig. 1(b), were used and connected to the VNA. The color scale from white to blue represents the intensity of the excited spin wave. White indicates a higher intensity of the excited spin wave. (b) Dependence of the spin wave responses on the magnitude of the magnetic field detected at Det. 3 when the magnetic field was 190 mT. (c) Comparison of the induced voltage under different exciters and the nonlinear effect of the interference detected at Det. 3. V(A), V(B), and V(A+B) represent the induced voltages when only Ext. A, only Ext. B, and both Ext. A and Ext. B, where this case corresponds to the simultaneous excitation, the spin-wave interference happened in the YIG medium. V(A) + V(B) corresponds to the linear combination of the individual excitations. The red line represents the difference between V(A+B) and V(A)+V(B), which corresponds to the nonlinear components of the interference. (d) Nonlinear component in the time window of 5.5–20 ns. (e) Waveform comparison across the different detectors: Det. 1, 3, 5, and 7 when the magnetic field was 182 mT. Diverse output waveforms with varying amplitudes and phases were also observed at the remaining detectors.

## B. Speech recognition task

In this study, we analyzed and evaluated four configurations (cases (i) to (iv)), which are the presence or absence of the cochleagram and interfered spin wave-based PRC, for two speech recognition tasks: spoken digit recognition and speaker classification. We assessed the impact of each combination on computational performance.

Figures 5(a) and (b) show the accuracies and confusion matrices in case (i), and the spoken-digit-recognition and speaker-classification accuracy were 85.8 % and 94.8 %, respectively. These results indicate that the frequency decomposition provided by cochleagram preprocessing significantly influences speech recognition tasks. Furthermore, they imply that comparable frequency decomposition capability is required in physical systems for PRC to handle such practical speech recognition tasks without cochleagram preprocessing. From the confusion matrix shown in Fig. 5(a), the number of correct predictions varies across digits in the spoken digit recognition task, suggesting that some digits cannot be correctly predicted using simple frequency decomposition alone. In contrast, in the speaker classification task, all predictions yielded correct results, with the number of correct predictions for each speaker exceeding 90. Next, Figures 5(c) and (d) show the accuracies and confusion matrices for each task in case (ii), where preprocessing cochleagram was used as input data and interfered spin wave-based PRC was also employed. This approach improved accuracy, achieving 95.2 % for the spoken digit recognition task and 97.0 % for the speaker classification task. Comparing the accuracy trends with case (i), the improvement is approximately 9 % for the spoken digit recognition task, whereas it is around 2 % for the speaker classification task. For both tasks, the frequency decomposition performed during preprocessing appears to have a significant impact, suggesting that the transformation and processing capabilities inherent in interfered spin wave-based PRC may not be fully utilized. Furthermore, the results for spoken digit recognition show an accuracy comparable to other PRCs (FeFET: 98.1 %,[48] optical circuit: 97 %,[5] spin torque oscillator: 99.8 %,[36] and so on. [6,26,34,40-42,45-47,49-51]). Examining the confusion matrix reveals that in Fig. 5(c), the number of correctly recognized digits for those frequently misrecognized in Fig. 5(a) improved. This indicates that the transformation capability provided by the interfered spin wave-based PRC influenced the recognition accuracy. However, in speaker classification, while the number of correct recognitions improved to some extent, the contribution of the interfered spin wave-based PRC is difficult to discern because all speakers originally recorded over 90 correct recognitions.

So far, we have evaluated the recognition accuracy of two types of speech recognition tasks using the cochleagram preprocessing method. However, the cochleagram requires substantial computational resources for preprocessing and is not well suited for edge AI applications. For low-power processing, it is desirable to perform speech recognition using the PRC alone. Therefore, we also evaluated the recognition performance without the cochleagram.

Figures 5(e) and (f) show the accuracies and confusion matrices of performing each task in case (iii). The results indicate that the system performed poorly, achieving only an 18.0 % accuracy for spoken digit recognition task and a 22.8 % accuracy for speaker classification task. These results confirm that these tasks cannot be solved using only linear processing, as the nonlinear transformation capability of the PRC is not utilized when only a linear readout is applied. Examining the confusion matrices for both tasks also reveals highly scattered prediction results, indicating that the tasks were barely solved at all. In contrast, speech recognition was performed using only the feature-extraction capabilities inherent to interfered spin wave-based PRC, i.e., case (iv). Figures 5(g) and (h) show the accuracies and confusion matrices for each task. These reveal significantly different accuracy: 29.0% for the spoken digit recognition task and 84.6% for the speaker classification task. This suggests differing computational demands across tasks, revealing that our setup demonstrates superior computational performance for speaker classification. The interfered spin wave-based PRC performs information processing that provides a frequency decomposition capability similar to that of the cochleagram, which appears to be sufficient for speaker classification. In addition, improved performance may be partly attributed to differences in how temporal information is utilized. Cochleagram-based preprocessing mainly emphasizes spectral information and doesn't fully utilize temporal structures present in speech signals due to separating into the time domains. In contrast, the PRC may exploit temporal dependencies in speech signals, which could contribute to improved performance. The confusion matrix in Fig. 5(f) shows that the number of correct predictions for each speaker varies significantly, ranging from 72 to 92. To achieve further performance improvement, a transformation that can capture the features of S3 currently being misrecognized is required. On the other hand, in the spoken digit recognition task shown in Fig. 5(h), an improvement in the number of correct predictions was observed for certain digits compared with the case (iii) in Fig. 5(e).

These results suggest that the nonlinear transformation capability of the nonlinear interfered spin wave-based PRC contributes to improved recognition performance. However, achieving high recognition accuracy for all digits requires more advanced processing capabilities, including frequency decomposition ability. Nevertheless, the present results demonstrate that speech recognition tasks performed without computationally expensive preprocessing such as the cochleagram, indicating the potential of nonlinear interfered spin wave-based PRC for practical edge-AI applications.

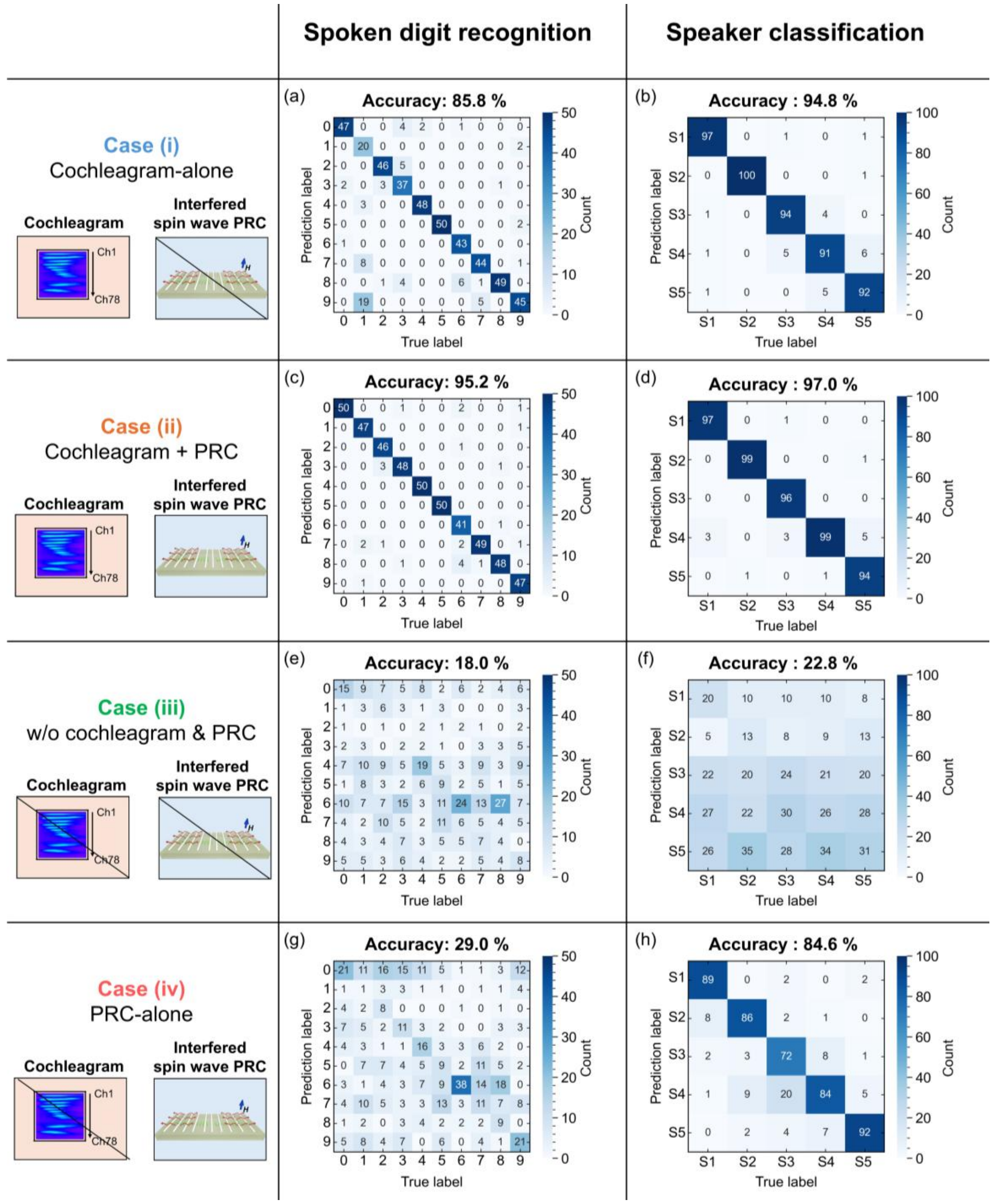

FIG. 5 (a,b) Accuracy and confusion matrices for case (i), where only cochleagram-preprocessed data were used, for (a) spoken digit recognition and (b) speaker classification. (c,d) Accuracy and confusion matrices for case (ii), where cochleagram-preprocessed data were used as the input to the interfered spin-wave-based PRC, for (a) spoken digit recognition and (b) speaker classification.

(e,f) Accuracy and confusion matrices for case (iii), where only input data obtained by converting raw audio signals into pulse sequences were used, in other words, corresponding to the case (iv) without interfered spin wave-based PRC processing, for (a) spoken digit recognition and (b) speaker classification. (g,h) Accuracy and confusion matrices for case (iv), where where only the interfered spin wave-based PRC was used, for (a) spoken digit recognition and (b) speaker classification

## CONCLUSION

In this study, accuracy of nonlinear interfered spin wave-based PRC was evaluated with the two types of speech recognition tasks, spoken digit recognition and speaker classification, which were conducted for four configurations defined by the presence or absence of the cochleagram and interfered spin wave-based PRC. When neither the cochleagram nor the interfered spin wave-based PRC was used, the accuracy exhibited low, approximately 20 %, confirming that linear processing alone cannot solve the speech recognition. Conversely, applying the cochleagram preprocessing raised accuracy to approximately 90 % in both tasks, indicating that frequency-decomposition capability is required for PRCs. When the preprocessed data were fed into the interfered spin-wave PRC, the accuracies reached 95.2 % for spoken digit recognition and 97.0 % for speaker classification. The modest gain over cochleagram-alone results suggests that the cochleagram dominates performance and that the transformation capability of interfered spin wave-based PRC is not fully exploited. When performing speech recognition using only the feature-extraction capability of the interfered spin wave-based PRC without the cochleagram, an accuracy of 85.8% was achieved in the speaker classification task. This suggests that the interfered spin wave-based PRC can handle practical tasks. On the other hand, in the spoken digit recognition task, the accuracy was around 29 %, showing only a slight improvement over the case using a linear processing alone, and the task remains challenging. This suggests that even for the same speech recognition task, the required processing capability differs depending on the classification target. By setting appropriate targets, it should be possible to construct a highly efficient speech recognition system that leverages strong nonlinearity and substantial short-term memory inherent in interfered spin wave-based PRC.

**Acknowledgments**

This work was in part supported by Innovative Science and Technology Initiative for Security Grant Number JPJ004596, ATLA, Japan. A part of this work was supported by “Advanced Research Infrastructure for Materials and Nanotechnology in Japan (ARIM)" of the Ministry of Education, Culture, Sports, Science and Technology (MEXT). Proposal Number JPMXP1225NM5244. A part of this work was supported by Japan Society for the Promotion of Science (JSPS) KAKENHI Grant Numbers JP25K17658.

**Author Declarations**

**Conflict of Interest**

The authors have no conflicts to disclose.

**Author contributions**

Sota Hikasa: Conceptualization; Investigation; Formal analysis; Writing–original draft; Writing–review & editing. Wataru Namiki: Conceptualization; Methodology; Investigation; Formal analysis; Writing–review & editing. Daiki Nishioka: Conceptualization; Formal analysis; Writing–review & editing. Maki Nishimura: Investigation. Ryo Iguchi: Conceptualization; Formal analysis; Writing–review & editing. Kazuya Terabe: Conceptualization; Supervision. Takashi Tsuchiya: Conceptualization; Formal analysis; Supervision; Writing–review & editing.

**Appendix A: Obtained diverse output waveforms**

The output waveforms obtained from each detector are shown in Fig. 6. The output waveforms exhibited diverse patterns due to different interference patterns of the obtained spin waves arising from the asymmetric layout and group velocity. These results indicate that the input data are nonlinearly transformed and mapped into a high-dimensional space.

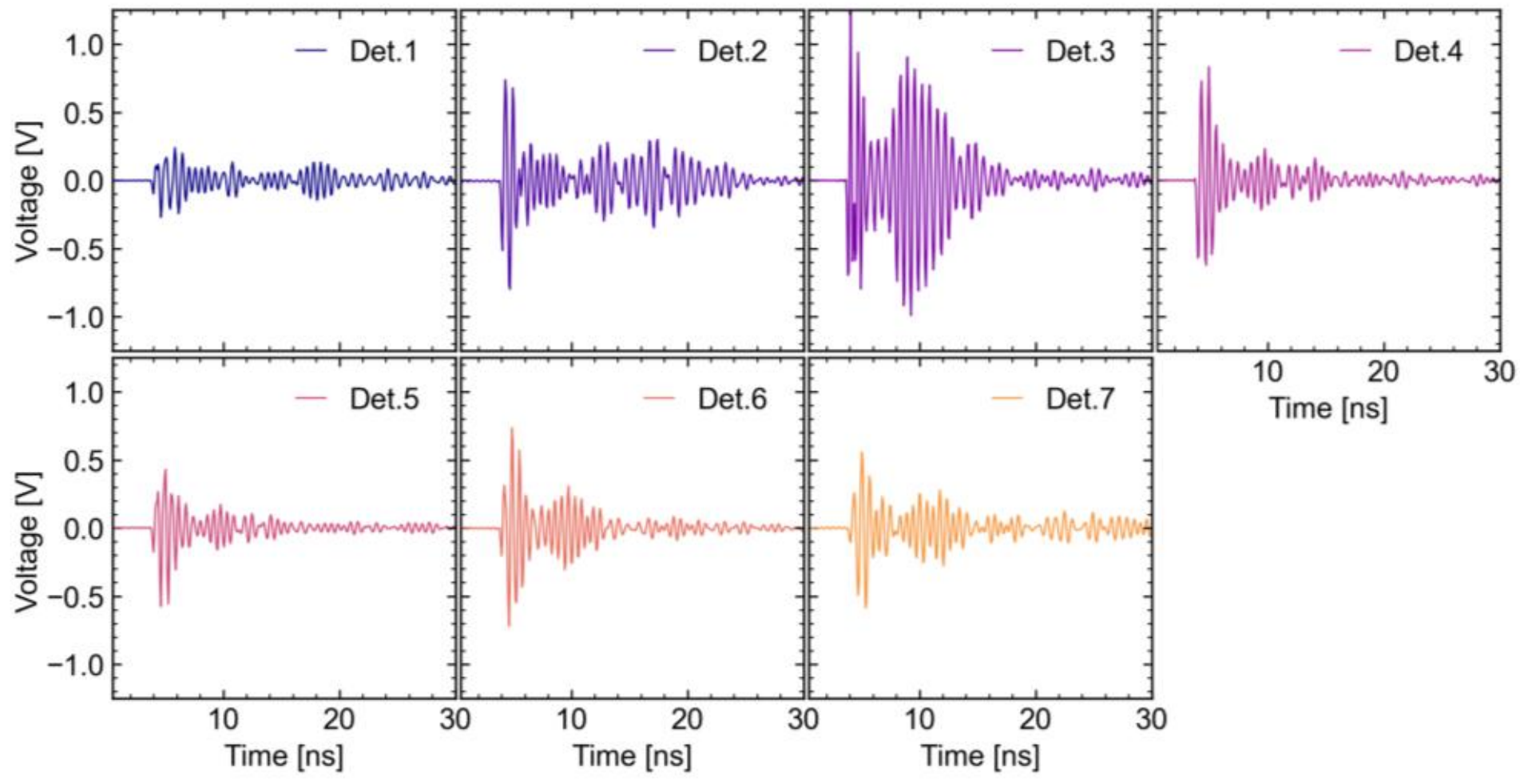

FIG. 6 Output waveforms obtained at each detector. The applied magnetic field was set to 182 mT.

**Appendix B: Optimization of the total number of nodes**

When addressing two types of speech recognition tasks using the interfered spin wave-based PRC, the total number of nodes was optimized. Figures 7(a) and (b) show the optimization results for each task in case (ii); i.e. using both cochleagram and interfered spin wave-based PRC. In both tasks, the recognition accuracy remains essentially unchanged while the number of virtual nodes $N_{\mathrm{v}}$ per detector is below 20. However, when the number of virtual nodes is further increased, the performance begins to decrease. In contrast, in case (iv), the total number of nodes $N$ is determined by $N_{\mathrm{v}} \times N_{\mathrm{pulse}} \times N_{\mathrm{det.}}$, varying and depending on two factors: the number of input pulses $N_{\mathrm{pulse}}$ contained in the time-domain window and the number of virtual nodes $N_{\mathrm{virtual\ node}}$ extracted from the output waveforms of each detector corresponding to the single input pulse. Therefore, both parameters were optimized. First, the optimization result with respect to the number of input pulses in the speaker classification task is shown in Fig. 7(c). The accuracy in the testing phase decreases significantly when the number of input pulses exceeds approximately 130. Comparing the total number of nodes and the number of time indices, it can be observed that the total number of nodes becomes comparable to or larger than the number of time indices when the number of input pluses exceeds 130. This means that the number of variables exceeds the number of equations during the weight optimization process, which results in overfitting. Next, the optimization result of the number of virtual nodes obtained by fixing the number of input pulses at 100 is shown in Fig. 7(d). In this case as well, the accuracy in the testing phase decreases when the total number of nodes becomes comparable to or larger than the number of time indices. This result also suggests that overfitting occurs for the same reason, leading to degraded performance.

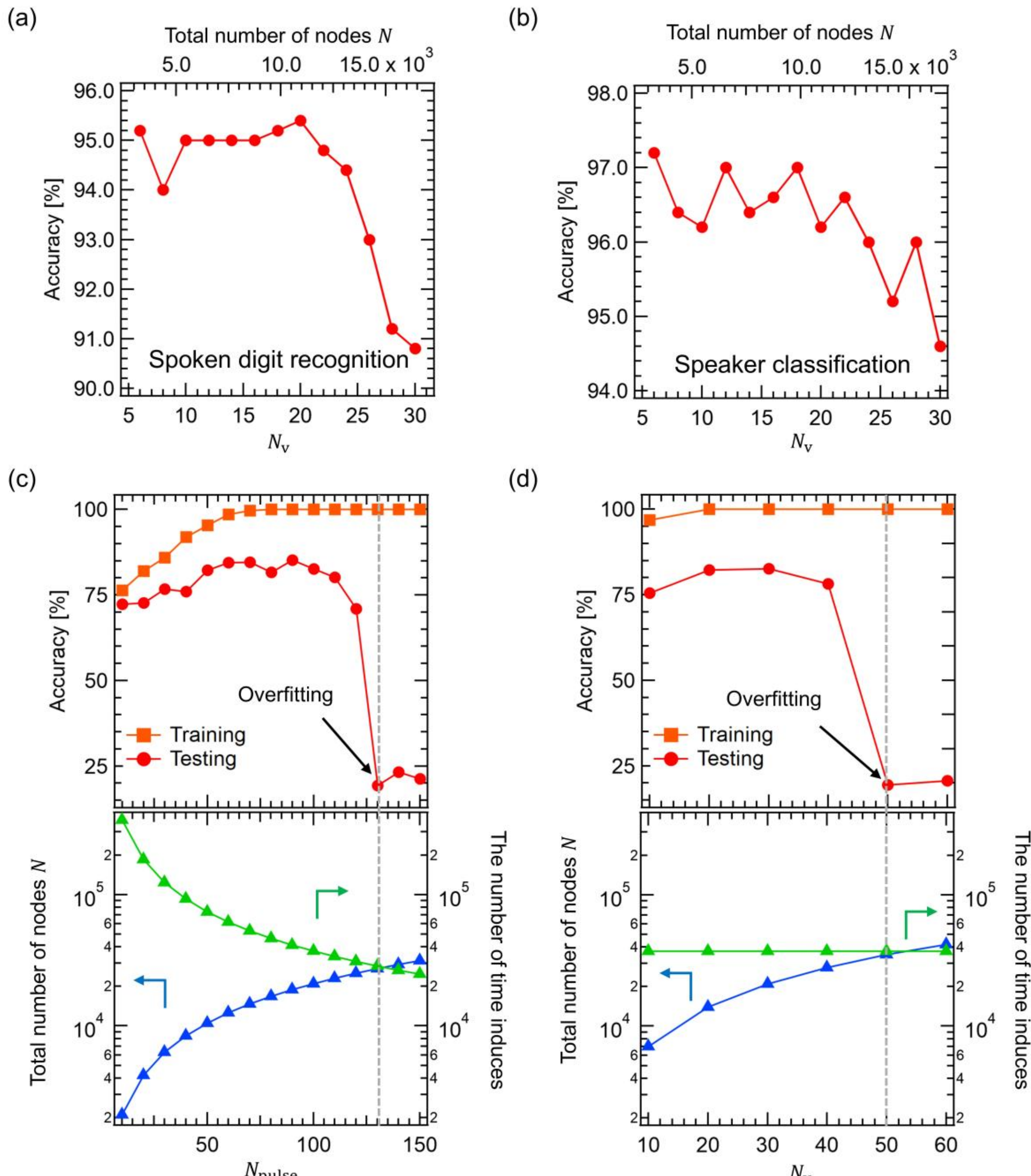

FIG. 7 (a,b) Accuracy dependence on the number of virtual nodes for (a) spoken digit recognition and (b) speaker classification when the magnetic field and pulse interval were 220 mT and 15 ns, respectively. The horizontal axis denotes the number of virtual nodes per detector, extracted from the output waveforms for a single input, defining as $N_{\mathrm{v}}$. (c) Dependence of (upper) classification accuracy and (lower) the total number of nodes $N$ and number of time indices on the number of input pulses for the speaker classification task without a cochleagram, when the magnetic field and pulse interval were 220 mT and 15 ns, respectively. The horizontal axis represents the number of input pulses $N_{\mathrm{pulse}}$ contained in the time-domain window, and the total number of nodes $N$ was defined by $N_{\mathrm{v}} \times N_{\mathrm{pulse}} \times N_{\mathrm{det.}}$. (d) Dependence of (upper) classification accuracy and (lower) the total number of nodes $N$ and number of time indices on the number of virtual nodes for the speaker classification task without a cochleagram, under the same experimental conditions (220 mT, 15 ns). The horizontal axis represents the number of virtual nodes, extracted from the output waveforms measured by each detector for a single input, defining as $N_{\mathrm{v}}$. Since the number of input pulses $N_{\mathrm{pulse}}$ was fixed at 100, the number of time indices was also fixed at 37144.